\documentclass[12pt]{article}
\usepackage{a4}
\begin{document}
\def\version{File W01v3c.tex last changed 15.01.01 by KK}
\def\nmonth{\ifcase\month\ \or January\or
   February\or March\or April\or May\or June\or July\or August\or
   September\or October\or November\else December\fi}
\def\nmonth{\ifcase\month\ \or January\or
   February\or March\or April\or May\or June\or July\or August\or
   September\or October\or November\else December\fi}
\def\rightheadline{\hfill\folio\hfill}
\def\leftheadline{\hfill\folio\hfill}
\def\operatorname#1{{\rm#1\,}}
\def\text#1{{\hbox{#1}}}
\def\birdy{\textstyle{d\over{d\epsilon}}|_{\epsilon=0}}
\def\qedbox{\hbox{$\rlap{$\sqcap$}\sqcup$}}
\def\pint{\operatorname{int}}\def\pext{\operatorname{ext}}
\def\cl{\operatorname{cl}}
\def\id{\text{I}}
\def\LL{{\mathcal{L}}}
\newcommand{\reals}{\mbox{${\rm I\!R }$}}
\newcommand{\nats}{\mbox{${\rm I\!N }$}}
\newcommand{\intgs}{\mbox{${\rm Z\!\!Z }$}}
\newcommand{\complex}{\mbox{${\rm C\!\!\!I}$}}

\newcommand{\nn}{\nonumber}
\newcommand{\iny}{\int_0^\infty\,\,dy\,\,}
\newcommand{\sump}{\sum_{p=0}^\infty}
\newcommand{\sumnu}{\sum_{\nu=0}^{m/2-1}}
\newcommand{\sumkn}{\sum_{k=0}^\infty}
\newcommand{\sumke}{\sum_{k=1}^\infty}
\newcommand{\suml}{\sum_{l=1}^\infty}
\newcommand{\enu}{e^{2\pi y -2\pi i \bar{\nu} /m}}
\newcommand{\nub}{\bar{\nu}}
\newcommand{\ang}{2\pi i \nub /m}

\newcommand{\laas}{(L_{aa}^+ +L_{aa}^-)}
\newcommand{\laaa}{(L_{aa}^+ -L_{aa}^-)}
\newcommand{\lbbs}{(L_{bb}^+ +L_{bb}^-)}
\newcommand{\lbba}{(L_{bb}^+ -L_{bb}^-)}
\newcommand{\lccs}{(L_{cc}^+ +L_{cc}^-)}
\newcommand{\lcca}{(L_{cc}^+ -L_{cc}^-)}

\newcommand{\labs}{(L_{ab}^+ +L_{ab}^-)}
\newcommand{\laba}{(L_{ab}^+ -L_{ab}^-)}
\newcommand{\lbcs}{(L_{bc}^+ +L_{bc}^-)}
\newcommand{\lbca}{(L_{bc}^+ -L_{bc}^-)}
\newcommand{\lcas}{(L_{ca}^+ +L_{ca}^-)}
\newcommand{\lcaa}{(L_{ca}^+ -L_{ca}^-)}
\newcommand{\lacs}{(L_{ac}^+ +L_{ac}^-)}
\newcommand{\laca}{(L_{ac}^+ -L_{ac}^-)}

\newcommand{\es}{(E^+ + E^-)}
\newcommand{\ea}{(E^+ - E^-)}

\newcommand{\rs}{(\tau^+ + \tau^-)}
\newcommand{\ra}{(\tau^+ - \tau^-)}

\newcommand{\rmms}{(R_{a\nu^+a\nu^+}+R_{a\nu^-a\nu^-})}
\newcommand{\rmma}{(R_{a\nu^+a\nu^+}-R_{a\nu^-a\nu^-})}

\newcommand{\rms}{(\tau_{;\nu^+}+\tau_{;\nu^-})}
\newcommand{\rma}{(\tau_{;\nu^+}-\tau_{;\nu^-})}

\newcommand{\ems}{(E_{;\nu^+}+E_{;\nu^-})}
\newcommand{\ema}{(E_{;\nu^+}-E_{;\nu^-})}

\newcommand{\rabcbs}{(R_{abcb}^+ + R_{abcb}^-)}
\newcommand{\rabcba}{(R_{abcb}^+ - R_{abcb}^-)}
\newcommand{\rambms}{(R_{a\nu^+b\nu^+}+R_{a\nu^-b\nu^-})}
\newcommand{\rambma}{(R_{a\nu^+b\nu^+}-R_{a\nu^-b\nu^-})}

\newcommand{\fms}{(f_{;\nu^+}+f_{;\nu^-})}
\newcommand{\fma}{(f_{;\nu^+}-f_{;\nu^-})}
\newcommand{\fmms}{(f_{;\nu^+\nu^+}+f_{;\nu^-\nu^-})}
\newcommand{\fmma}{(f_{;\nu^+\nu^+}-f_{;\nu^-\nu^-})}

\newcommand{\m}{l+(m-2)/2}

\newtheorem{assumption}{Assumption}[section]
\newtheorem{theorem}[assumption]{Theorem}
\newtheorem{conjecture}[assumption]{Conjecture}
\newtheorem{lemma}[assumption]{Lemma}
\newtheorem{definition}[assumption]{Definition}
\def\trace{\operatorname{Tr}}
\def\bork{\newline\phantom{.}\qquad}
\def\BB{{\mathcal{B}}}
\def\dvxi{U}
\title{Heat trace asymptotics with transmittal boundary
       conditions and quantum brane-world scenario
}
\author{Peter B. Gilkey$^a$\thanks{EMAIL: gilkey@darkwing.uoregon.edu}
,
     Klaus Kirsten$^b$\thanks{EMAIL: klaus@a35.ph.man.ac.uk}
 and Dmitri V. Vassilevich$^c$\thanks{EMAIL:
vassil@itp.uni-leipzig.de On leave from V.A. Fock Department of
Theoretical Physics, St.Petersburg University, 198904 Russia}
\\[5pt]
{\it $^a$Department of Mathematics, University of
Oregon,}\\{\it Eugene OR 97403 USA. }\\
$^b${\it Department of Physics and Astronomy, 
    The University of
}\\{\it     Manchester, Oxford Road, Manchester UK M13 9PL UK
     } \\
$^c${\it Institute for Theoretical Physics, 
University of Leipzig,}\\{\it Augustusplatz 10, 04109 Leipzig, Germany. 
}}
\maketitle

\begin{abstract}
We study the spectral geometry of an operator of Laplace type
on a manifold with a singular surface. We calculate several
first coefficients of the heat kernel expansion. These
coefficients are responsible for divergences 
and conformal anomaly in quantum
brane-world scenario. 
\end{abstract}

\section{Motivations}
It is well known that the regularized
one-loop effective action in Euclidean
quantum field theory is given by the following formal expression
\begin{equation}
W^{\mbox{\scriptsize reg}}=\frac 12 \log \det (D)_{\mbox{\scriptsize 
reg}}
= -\frac{\mu^{2s}}{2}
\int_0^\infty dt~t^{s-1}
{\rm Tr} (\exp (-tD)) \,, \label{effact}
\end{equation}
where we have introduced the (zeta-function) regularization
parameter $s$, which should be set to zero after calculations.
The parameter $\mu$ of the dimension of mass makes the effective
action (\ref{effact}) dmensionless for any $s$. The value of $\mu$
is to be fixed by a normalization condition.
The operator $D$ is a partial
differential operator which appears in the quadratic part
of the classical action. We assume that $D$ is a second order
operator of Laplace type and that there is an asymptotic series
\begin{equation}
{\rm Tr}(f\exp (-tD))
\cong \sum_{n=0}^\infty t^{\frac{n-m}2} a_n (f,D)
\label{hkD}
\end{equation}
as $t\downarrow 0$. Here $m$ is the dimension of the underlying manifold 
$M$ and $f$ is a smearing (or localizing) function. Near $s=0$ the regularized
effective action behaves as
\begin{equation}
W^{\mbox{\scriptsize reg}}\cong -\frac 1{2s} a_m(1,D)+O(s^0)\,.
\label{div}
\end{equation}
Therefore, the heat kernel coefficient $a_m$ provides complete
information on the one-loop divergences. In most of the cases
that one considers,
the coefficients $a_n$ are locally computable; equivalently, this means
that the counter-terms are local. If the operator $D$ is conformally
covariant, then $a_m$ also defines the trace anomaly in the stress-energy
tensor.

The heat kernel asymptotics on (smooth) manifolds with or
without a boundary have been studied in some detail. Relatively
less is known about the case when there are some kinds of
``non-smoothness'' inside the manifold. Only the cases of
point-like singularities, either conical 
\cite{Ch87,BS91,Cognola:1994qg,Fursaev:1997uz,Dowker:1994bj} or
delta-function ones \cite{AGHH88}, have attracted considerable
attention. We also mention a related work \cite{Solodukhin:1999xn}.

In the present paper we deal with the heat kernel asymptotics
for the case when the operator $D$ has a ``non-smoothness''
on a surface $\Sigma$ of co-dimension one. Such kind of singularities
appear in many problems of quantum field theory as, e.g. the Casimir
energy calculations. The case when the metric is smooth across 
$\Sigma$ has been studied recently by Bordag and Vassilevich
\cite{BV99} and by Moss \cite{Moss00}. In the present paper
we allow normal derivatives of the metric to jump on
$\Sigma$. This study is motivated by (and has applications in)
the brane-world scenario \cite{RS99,Rubakov:1983bb} 
which operates with the metric of the type
\begin{equation}
(ds)^2=(dx^5)^2+e^{-\alpha |x^5|} (ds_4)^2 \,,\label{bwmet}
\end{equation}
where $\alpha$ is a constant and where $(ds_4)^2$ is a line element
on four-dimensional hypersurface. Due to the presence of
the absolute value of the 5th coordinate in (\ref{bwmet}), the
normal derivative of the metric jumps on the surface $\Sigma$ defined
by the vanishing of the coordinate $x^5$.
It is also assumed that the bulk action is supplemented
by a surface term concentrated on $\Sigma$. This model can be further
generalized to allow for a more general line element and a more
general singular surface $\Sigma$. One can also imagine
a similar construction in dimension $m$ other than $5$, though
the codimension of $\Sigma$ will be always supposed to be $1$.
It is clear that
the quadratic part of the classical matter action for a quite general 
class
of the brane-world models should be of the form
\begin{equation}
S_2=\int_M d^5x\sqrt{g} \phi D \phi \,,\label{S2}
\end{equation}
where $\phi$ describes the bulk field fluctuations, and
the operator $D$ is\footnote{Note that in the present
paper we neglect possible derivative terms in the surface action
for simplicity}
\begin{equation}
D=-(\nabla^2 +E(x))+\dvxi \delta_\Sigma \,.\label{singop}
\end{equation}
Here $\nabla$ is a suitable covariant derivative, and $E(x)$ and $\dvxi 
(x)$
are endomorphisms (matrix valued fields). Let $h$ be
the determinant of the induced metric on $\Sigma$. Then $\delta_\Sigma$ is a
delta function defined such that
\begin{equation}
\int_M dx\sqrt{g} \delta_\Sigma f(x) =\int_\Sigma dx \sqrt{h}
f(x) \,.\label{delta}
\end{equation}

We shall assume that $D$ is smooth on $M-\Sigma$. On the hypersurface
$\Sigma$, we shall only assume that the leading symbol (metric) of
$D$ is continuous; the normal derivatives of the metric are not assumed to be
continuous on $\Sigma$. Furthermore, 
we shall impose no assumption of continuity
on the remaining tensors ($E$, curvature, etc.) on $\Sigma$. 

Let $x^m$ be a smooth function so the equation $x^m=0$ defines the
hypersurface $\Sigma$ and so $dx^m\ne0$ on $\Sigma$. It is convenient to
introduce a coordinate system on
$M$ such that in a neighbourhood of $\Sigma$
\begin{equation}
(ds)^2=(dx^m)^2+ g_{ab}dx^adx^x. \label{mmet}
\end{equation}

The spectral problem
for $D$ on $M$ as it stands is ill-defined owing to the discontinuities (or
singularities) on $\Sigma$. It should be replaced by a pair of spectral
problems on the two sides
$M^\pm$ of
$\Sigma$ together with suitable matching conditions on $\Sigma$. In order to
find such matching conditions, we 
consider an eigenfunction $\phi_\lambda$ of the operator 
(\ref{singop}):
\begin{equation}
D\phi_\lambda =\lambda \phi_\lambda \,.
\label{eigen}
\end{equation}
It is clear that $\phi_\lambda$ must be continuous on $\Sigma$:
\begin{equation}
\phi\vert_{x^m=+0}=\phi\vert_{x^m=-0} \,.\label{mc1}
\end{equation}
Otherwise, the second normal derivative of $\phi_\lambda$ would create
a $\delta'$ singularity on $\Sigma$ which is absent on the right
hand side of (\ref{eigen}). Let us integrate (\ref{eigen}) over
a small cylinder ${\cal C}=C^{m-1}\times [-\epsilon ,+\epsilon]$
\begin{equation}
\int_{\cal C} d^mx\sqrt{g} \left( -\nabla_m^2 \phi_\lambda
-\left[ \nabla_a^2 \phi_\lambda +(E+\lambda)\phi_\lambda \right]\right)
+\int_C d^{m-1}x\sqrt h \dvxi\phi_\lambda =0\,. \label{intcyl}
\end{equation} 
We now take the limit as $\epsilon\to 0$. Since the expression in the
square brackets in (\ref{intcyl}) is bounded, the contribution
that this term makes vanishes in the limit. We obtain
\begin{equation}
0=\int_C d^{m-1}x\sqrt h \left( -\nabla_m \phi_\lambda\vert_{x^m=+0}
+\nabla_m \phi_\lambda\vert_{x^m=-0} +\dvxi\phi_\lambda \right)\,.
\label{int2cyl}
\end{equation}
Since $C$ and $\lambda$ are arbitrary, we conclude that a proper
matching condition for the normal derivatives is
\begin{equation}
-\nabla_m \phi \vert_{x^m=+0}
+\nabla_m \phi \vert_{x^m=-0} +\dvxi\phi =0 \,.\label{dermatch}
\end{equation}
A more mathematically careful construction of these transmittal
boundary conditions will be given in subsequent sections.

There have been already many works devoted to the quantization of 
bulk fields\footnote{Not to be mixed with quantum effects of
the so-called ``brane matter'' which is confined on the singular
surface} in
the brane-world scenario (see e.g. 
\cite{Garriga:2000jb,Toms:2000vm,Forste:2000ft,
Goldberger:2000dv,Nojiri:2000bz,Alvarez:2000vr,Brevik:2000vt,Hofmann:2000cj}).
However, the heat kernel expansion, divergences and renormalization
have not been discussed to a considerable order of generality.

Here is a brief guide to this paper; a more expanded discussion is given in
Section 2 after the necessary notation has been introduced. In
Section 2, we give a more precise statement of transmittal boundary conditions
and discuss the geometry of operators of Laplace type. In section 3 we
consider a smooth structure and the gluing construction. The invariance theory
is developed in section 4. Section 5 deals with reduction of the transmittal
problem to Dirichlet and Neumann boundary value problems. In section 6, we
construct a transmittal problem for the de Rham complex. We use this
problem to complete the calculation of several first heat kernel coefficients.
The coefficient $a_4$ is calculated in section 7. In section 8 we calculate
$a_5$ for a restricted class of transmittal problems and discuss applications
to the brane-world scenario. Appendix contains some technical details.

\section{Introduction}
 Let $\Sigma$ be a codimension $1$ hypersurface of a compact smooth 
manifold
which divides $M$ into two manifolds $M^\pm$. This means that
$$M:=M^+\cup_\Sigma M^-$$
is the union of two compact manifolds $M^\pm$ along their common
boundary $\Sigma$. We assume given a Riemannian metric which is 
continuous on
$M$ and smooth when restricted to $M^\pm$. Let $V$ be a smooth vector 
bundle
over
$M$ and let
$D^\pm$ be operators of Laplace type on
$V^\pm:=V|_{M^\pm}$; no further conditions are placed on $D^\pm$ apart 
from the
assumption that the leading symbols agree on $\Sigma$. The operators 
$D^\pm$
determine canonical connections ${}^\pm\nabla$ on $V^\pm$, see equation
(\ref{arefac}) below. Let $\dvxi$ be an auxiliary endomorphism of
$V_\Sigma:=V|_\Sigma$. Let $\nu$ be the inward unit normal of 
$\Sigma\subset
M^+$ and let
$\phi:=(\phi^+,\phi^-)$ be a pair of smooth sections to $V^\pm$.
We define the {\it transmittal operator}
\begin{equation}\label{arefaa}
  {\mathcal{B}}_\dvxi\phi=\{\phi^+|_\Sigma-\phi^-|_\Sigma\}\oplus
  \{(\nabla^+_\nu\phi^+)|_\Sigma
   -(\nabla^-_\nu\phi^-)|_\Sigma-\dvxi\phi^+|_\Sigma\}.
\end{equation}
An elliptic boundary condition for a $q^{th}$ order operator on a 
vector
bundle of dimension $r$ must involve $\frac12qr$ conditions. We set 
$q=2$ as we
are considering operators of Laplace type. Neumann and Dirichlet 
boundary
conditions involve
$\frac122r=r$ conditions. Transmittal boundary conditions fulfil this 
counting
condition; since we have two vector bundles
$V^\pm$, we must specify
$\frac122(2r)=2r$ conditions which is what the vanishing of the 
operator
in equation (\ref{arefaa}) imposes:
$$\phi^+|_\Sigma=\phi^-|_\Sigma\text{ and }
  \nabla_\nu^+\phi^+|_\Sigma=\{\nabla_\nu^-\phi^-|_\Sigma\}+\dvxi\{\phi^+|_\Sigma\}.$$

Let $D:=(D^+,D^-)$ act on $\phi:=(\phi^+,\phi^-)$ in the natural 
fashion. We
restrict the domain of $D$ to pairs $\phi$ so that 
${\mathcal{B}}_\dvxi\phi=0$.
Let
$D_{{\mathcal{B}_\dvxi}}$ be the realization of $D$ on this domain and 
let
$e^{-tD_{{\mathcal{B}}_\dvxi}}$ be the associated fundamental solution 
of the
heat equation. Let
$f=(f^+,f^-)$ where the
$f^\pm$ are smooth on
$M^\pm$ and where $f^+|_\Sigma=f^-|_\Sigma$; no matching is assumed for 
the
normal derivatives of $f$. Let
$$a(f,D,\dvxi)(t):=\trace_{L^2}\{fe^{-tD_{{\mathcal{B}}_\dvxi}}\}$$
be the heat trace. If the $D^\pm$ are formally self-adjoint, and if 
$\dvxi$ is
self-adjoint, then
$D_{\mathcal{B}_\dvxi}$ self-adjoint. Thus we can find a discrete 
spectral
resolution
$\{\lambda_i,\phi_i\}$ where the
$\{\phi_i\}$ form a complete orthonormal basis for $L^2(V)$, where
$D^\pm\phi_i^\pm=\lambda_i\phi_i^\pm$, and where
  ${\mathcal{B}}_\dvxi\phi=0$.
We then have:
\begin{equation}\label{arefab}
a(f,D,\dvxi)(t)
  =\textstyle\sum_ie^{-t\lambda_i}
  \textstyle\int_Mf(\phi_i,\phi_i).
\end{equation}

\begin{assumption}\label{refass21} There exists a full asymptotic series as 
$t\downarrow0$:
$$a(f,D,\dvxi)(t)
     \sim\textstyle\sum_{n\ge0}t^{(n-m)/2}a_n(f,D,\dvxi)$$
where the heat trace coefficients $a_n(f,D,\dvxi)$ are locally
computable, i.e. there are local invariants $a_n(x^\pm,D^\pm)$ defined 
on
$M^\pm$ and local invariants $a_n^\Sigma(y,f,D,\dvxi)$ defined on
$\Sigma$ so that:
\begin{eqnarray*}
  &&a_n(f,D,\dvxi)=a_n^+(f,D)+a_n^-(f,D)+a_n^\Sigma(f,D,\dvxi)\text{ 
where}\\
  &&a_n^\pm(f,D)=\textstyle\int_{M^\pm}f(x^\pm)a_n(x^\pm,D^\pm)\text{ 
and}\\
  &&a_n^\Sigma(f,D,\dvxi)=\textstyle\int_\Sigma
a_n^\Sigma(y,f,D,\dvxi).\end{eqnarray*}\end{assumption} 
We remark that Assumption \ref{refass21} has been established by  \cite{BV99,Moss00}
if the leading symbol (i.e. the metric) is smooth.

Before discussing the interior invariants $a_n^\pm$, we must describe 
the
geometry of operators of Laplace type. The operators
$D^\pm$ determine natural connections $\nabla^\pm$ and natural $0^{th}$ 
order
operators $E^\pm$ so that
$$D^\pm=-\{\trace(\nabla^\pm\nabla^\pm)+E^\pm\}.$$
If we choose a system of local coordinates and a local frame, we can 
express:
$$D^\pm=-(g^{\pm,\mu\nu}\partial_\mu\partial_\nu
   +A^{\pm,\mu}\partial_\mu+B^\pm)$$
where we adopt the Einstein convention and sum over repeated indices. 
Let
$\Gamma^\pm$ be the Christoffel symbols of the metrics
$g^\pm$. The connection
$1$ forms
$\omega^\pm$ of
$\nabla^\pm$ and the endomorphisms
$E^\pm$ are then given by
\begin{eqnarray}
   &&\textstyle\omega^\pm_\delta=\frac12g_{\nu\delta}^\pm
     (A^{\pm,\nu}+g^{\pm,\mu\sigma} 
    \Gamma_{\mu\sigma}^\pm{}^\nu I)\text{ and}\nonumber\\
   &&{}E^\pm=B^\pm-g^{\pm,\nu\mu}(\partial_\nu{}\omega_\mu^\pm
    +\omega^\pm_\nu\omega^\pm_\mu
    -\omega^\pm_\sigma\Gamma_{\nu\mu}^\pm{}^\sigma);
    \label{arefac}\end{eqnarray}
see \cite{G94} for further details. Let indices $i$, $j$, $k$, and $l$
range from $1$ to $m$ and index a local orthonormal frame for the 
tangent
bundle of the manifold. Let
$R_{ijkl}^\pm$ be the components of the curvature tensor of the 
Levi-Civita
connection; with our sign convention the Ricci tensors $\rho^\pm$ and 
the scalar
curvatures $\tau^\pm$ are given by:
$$\rho_{ij}^\pm:=R_{ikkj}^\pm\text{ and 
}\tau^\pm:=\rho_{ii}=R_{ijji}.$$
Let $\Omega_{ij}^\pm$ be the components of the curvature tensors of the
connection $\nabla^\pm$. The interior invariants have been computed
previously in the smooth context. They vanish if $n$ is odd and have 
been
determined explicitly for $n=0,2,4,6,8,10$ - see for example 
\cite{ABC89, A90,G75,vandeVen:1998pf}. The presence of the junction 
discontinuity along $\Sigma$ does not affect
the interior invariants
$a_n^\pm$ and consequently we may apply these results to see that:

\begin{theorem}\label{Theoremanm} The invariants
$a_n^\pm$ vanish if $n$ is odd. We have:\begin{enumerate}
\smallskip\item $a_0^\pm(f,D)=(4\pi)^{-m/2}\int_{M^\pm}f\trace(I)$.
\smallskip\item $a_2^\pm(f,D)=(4\pi)^{-m/2}\textstyle\frac16
     \int_{M^\pm}f\trace(\tau^\pm I+6E^\pm)$.
\smallskip\item $a_4^\pm(f,D)=(4\pi)^{-m/2}\textstyle\frac1{360}
    \int_{M^\pm}f\trace\{60E^\pm_{;kk}+60R^\pm _{ijji}E^\pm
    +180E^\pm E^\pm$\bork$
   +30\Omega^\pm_{ij}\Omega^\pm_{ij}
    +(12\tau^\pm_{;kk}+5(\tau^\pm)^2-2|(\rho^\pm)^2|
     +2|(R^\pm)^2|)I\}$.
\end{enumerate}\end{theorem}

We now introduce some additional notation to describe the invariants
$a_n^\Sigma$. Let indices $a$, $b$, $c$, and $d$ index a local 
orthonormal
frame $\{e_a\}$ for the tangent bundle of $\Sigma$; we complete this 
frame to a
frame for the tangent bundle of $M$ by letting $e_m:=\nu$ be the inward 
unit
normal of $\Sigma\subset M^+$. Let $\nu^\pm:=\pm\nu$ be the inward unit
normals of $\Sigma\subset M^\pm$ and let
$$L_{ab}^\pm:=(\nabla_{e_a}^\pm e_b,\nu^\pm)|_\Sigma$$ 
be the associated second fundamental forms. Let
$$\omega_a:=\nabla_a^+-\nabla_a^-.$$
Since the difference of two connections is tensorial, $\omega_a$ is a
well defined endomorphism of $V_\Sigma$. The tensor
$\omega_a$ is {\it chiral}; it changes sign if the roles of $+$ and $-$ 
are
reversed. Since we can describe the matching condition on the normal
derivatives in the form:
$$(\nabla^+_{\nu^+}\phi^+)|_\Sigma+(\nabla^-_{\nu^-}\phi^-)|_\Sigma=
   \dvxi\phi|_\Sigma,$$
the tensor field
$\dvxi$ is {\it non-chiral} as it is not sensitive to the roles of $+$ 
and $-$.
The main result of this paper is the following Theorem which determines 
the
invariants
$a_n^\Sigma$ for
$n=0,1,2,3$; the invariant
$a_4^\Sigma$ is a bit more combinatorially complex and the formula for 
this
invariant is discussed in Section \ref{SectA4}.

\begin{theorem}\label{ThmA4Sigma}\ 
\begin{enumerate}\item $a_0^\Sigma(f,D,\dvxi)=0$.
\item $a_1^\Sigma(f,D,\dvxi)=0$.
\item $a_2^\Sigma(f,D,\dvxi)=
     \textstyle(4\pi)^{-m/2}\frac16\textstyle\int_\Sigma
       \trace\{2f(L_{aa}^++L_{aa}^-)I-6f\dvxi\}$.
\item $a_3^\Sigma(f,D,\dvxi)=\textstyle(4\pi)^{(1-m)/2}\frac1{384}
\textstyle\int_\Sigma\trace\{\frac32f(L_{aa}^+L_{bb}^++L_{aa}^-L_{bb}^-+2L_{aa}^+L_{bb}^-)I
$\par\qquad$
    +3f(L_{ab}^+L_{ab}^++L_{ab}^-L_{ab}^-+2L_{ab}^+L_{ab}^-)I
    +9(L_{aa}^++L_{aa}^-)(f_{;\nu^+}^++f_{;\nu^-}^-)I
$\par\qquad$
+48f\dvxi^2+24f\omega_a\omega_a
    -24f(L_{aa}^++L_{aa}^-)\dvxi
    -24(f_{;\nu^+}^++f_{;\nu^-}^-)\dvxi\}$.
\end{enumerate}\end{theorem}
We can now give a more complete outline to the paper than was given in the
introduction. In Section \ref{SectSmooth}, we
give an alternate formulation of transmittal boundary conditions in 
terms of
$C^1$ structures that will be convenient when considering conformal 
variations.
In Section
\ref{SectInvar}, we use invariance theory and dimensional analysis to 
prove
the following result which gives the general form that the invariants
$a_n^\Sigma$ have:

\begin{lemma}\label{arefc} There exist universal constants so that:
\begin{enumerate}
\item $a_0^\Sigma(f,D,\dvxi)=0$.
\item $a_1^\Sigma(f,D,\dvxi)=\textstyle\int_\Sigma c_1f\trace(I)$
\item 
$a_2^\Sigma(f,D,\dvxi)=\textstyle(4\pi)^{-m/2}\frac16\textstyle\int_\Sigma
\trace\{d_1f(L_{aa}^++L_{aa}^-)I+d_2(f_{;\nu^+}^++f_{;\nu^-}^-)I+
   d_3f\dvxi\}$.
\item $a_3^\Sigma(f,D,\dvxi)=\textstyle(4\pi)^{(1-m)/2}\frac1{384}
\textstyle\int_\Sigma\trace\{c_2(L_{aa}^+L_{bb}^-)I
     +c_3(L_{ab}^+L_{ab}^-)I
$\par\qquad$
    +c_4(L_{aa}^+-L_{aa}^-)(f_{;\nu^+}^+-f_{;\nu^-}^-)I
    +c_5(f_{;\nu^+\nu^+}^++f_{;\nu^-\nu^-}^-)I
$\bork$
       +c_6(E^++E^-)
    +c_7(R_{ijji}^++R_{ijji}^-)I+c_8(\rho_{mm}^++\rho_{mm}^-)I
$\bork$
    +d_4f(L_{aa}^+L_{bb}^++L_{aa}^-L_{bb}^-+2L_{aa}^+L_{bb}^-)I
    +d_5f(L_{ab}^+L_{ab}^++L_{ab}^-L_{ab}^-+2L_{ab}^+L_{ab}^-)I
$\bork$
    +d_6(L_{aa}^++L_{aa}^-)(f_{;\nu^+}^++f_{;\nu^-}^-)I
+d_7f\dvxi^2
    +d_8f(L_{aa}^++L_{aa}^-)\dvxi$\bork$
   +d_9(f_{;\nu^+}^++f_{;\nu^-}^-)\dvxi+e_1f\omega_a\omega_a\}$.
\end{enumerate}\end{lemma}

If we suppose that the operator $D$ is smooth and that the localizing function $f$ is
smooth on all of  $M$, then $\Sigma$ plays no  role and thus the invariants $a_n^\Sigma$
vanish. We use this observation to  show in
Lemma \ref{crefb} that the coefficients $c_i$ must vanish. In Section
\ref{SectBoundary} we recall formulas for the heat trace invariants on
manifolds with boundary; see Lemma \ref{drefa}. We use these formulas 
to
determine the coefficients
$d_j$, see Lemma \ref{drefc} for details. In Section \ref{SectDeRham}, we
construct a transmittal problem for the de Rham complex and use the 
resulting
local index theorem to show that the one remaining unknown coefficient 
has the
value $e_1=24$; this completes the proof of Theorem
\ref{ThmA4Sigma}.  We remark that Moss \cite{Moss00} used different 
methods to
show that $e_1=24$. In Section
\ref{SectA4}, we perform a similar analysis to determine the invariant
$a_4^\Sigma$. 
The value of the coefficients $c_1$,
$d_1$, $d_2$, $d_3$, $c_5$, $c_6$, $c_7$, and
$c_8$ agrees with the values calculated previously in \cite{BV99} using 
other
methods.

\section{Glueing constructions}\label{SectSmooth}

We use the geodesic flow to identify a neighborhood of $\Sigma$ in 
$M^+$ with
$\Sigma\times[0,\varepsilon)$ and a neighborhood of $\Sigma$ in $M^-$ 
with
$\Sigma\times(-\varepsilon,0]$ for some $\varepsilon>0$ so that the 
curves
$t\rightarrow(y,t)$ are unit speed geodesics normal to the boundary
$\Sigma:=\Sigma\times\{0\}$. We define a canonical smooth structure on
$M=M^+\cup M^-$ by glueing along $\Sigma\times\{0\}$. Note that the 
metric
then takes the form:
$${}^\pm ds^2=g_{ab}^\pm(y,t)dy^a\circ dy^b+dt\circ dt.$$
We can use $\dvxi$ to define a canonical $C^1$ structure on $V$. Let
$s_\Sigma$ be a local frame for $V|_\Sigma$. We use parallel transport 
along
the geodesic normals to define a local frame $s^-$ for $V^-$ near 
$\Sigma$ so
$\nabla_\nu s^-=0$. We twist a corresponding parallel frame over $V^+$ 
to
define a local frame $s^+$ for $V^+$ near $\Sigma$ so $\nabla_\nu 
s^+=\dvxi s^+$.
We glue $s^+$ to $s^-$ over $\Sigma$ to define a $C^1$ structure for 
$V$ over
$M$ which is characterized by the property that
$\nabla^+_\nu-\nabla^-_\nu=\dvxi$. We then have that
${\mathcal{B}}_\dvxi\phi=0$ if and only if $\phi\in C^1(V)$. When 
studying
variations of the form $D(\varepsilon):=e^{\varepsilon f}D$ we will fix 
the
$C^1$ structure or equivalently choose $\dvxi(\varepsilon)$ so the 
transmittal
boundary condition
${\mathcal{B}}_\dvxi(\varepsilon)$ is independent of $\varepsilon$.

Suppose that the bundles $V^\pm$ are equipped with Hermitian inner 
products and
that the operators $D^\pm$ are formally self-adjoint. This means that 
the
associated connections $\nabla^\pm$ are unitary and the endomorphisms 
$E^\pm$
are symmetric. Suppose that $\dvxi$ is self-adjoint. Let 
$\phi:=(\phi^+,\phi^-)$
and $\psi:=(\psi^+,\psi^-)$ satisfy transmittal boundary conditions. 
Since $\nu$
is the inward unit normal of
$\Sigma\subset M^+$ and the outward unit normal of $\Sigma\subset M^-$, 
we may
integrate by parts to show that $D$ is self-adjoint by computing:
\begin{eqnarray}
&&(D\phi,\psi)_{L^2}-(\phi,D\psi)_{L^2}\nonumber\\
&=&\textstyle\int_{M^+}\{(\phi^+_{;ii},\psi^+)-(\phi^+,\psi_{;ii}^+)\}
   +\textstyle\int_{M^-}\{(\phi^-_{;ii},\psi^-)-(\phi^-,\psi_{;ii}^-)\}
   \nonumber\\
&=&\textstyle\int_{M^+}\{(\phi^+_{;i},\psi^+)-(\phi^+,\psi_{;i}^+)\}_{;i}
   +\textstyle\int_{M^-}\{(\phi^-_{;i},\psi^-)-(\phi^-,\psi_{;i}^-)\}_{;i}
   \label{brefaa}\\
&=&-\textstyle\int_\Sigma\{(\phi^+_{;\nu}-\phi^-_{;\nu},\psi)
    -(\phi,\psi^+_{;\nu}-\psi^-_{;\nu})\}\nonumber\\
&=&-\textstyle\int_\Sigma\{(\dvxi\phi,\psi)-(\phi,\dvxi\psi)\}=0.\nonumber
\end{eqnarray}

\section{Invariance Theory}\label{SectInvar}

We begin by giving the proof of Lemma \ref{arefc}. We assign degree $1$ 
to the
tensors
$\{L^\pm,\dvxi,\omega\}$ and assign degree $2$ to the tensors
$\{R^\pm,\Omega^\pm,E^\pm\}$. We increment the degree by $1$ for
every explicit covariant derivative which appears. Dimensional analysis 
shows
that the integrands
$a_n^\Sigma$ can be built universally and polynomial from monomials 
which are
homogeneous of weighted degree $n-1$ and which are non-chiral. The 
structure
group is
$O(m-1)$. We use H. Weyl's theorem on the invariants of the orthogonal 
group
to write down a spanning set; product formulas then yield the 
coefficients are
dimension free except for the normalizing factor of
$(4\pi)^{-m/2}$. \qedbox

Thus to determine the formulas for the $a_n^\Sigma$, we must determine 
the unknown
coefficients in Lemma \ref{arefc}. We shall use the various functorial 
properties of
these invariants in the calculation. We begin our evaluation with:

\begin{lemma}\label{crefb} We have $c_1=c_2=c_3=c_4=c_5=c_6=c_7=c_8=0$.
\end{lemma}

\noindent{\bf Proof:} Suppose we take $\dvxi=0$ and let $(f,D)$ be 
smooth on
all of
$M$. Then the hypersurface $\Sigma$ plays no role and thus the 
invariants
$a_n^\Sigma$ vanish in this setting. The terms indexed by these 
coefficients
$c_1$, $c_2$, $c_3$, $c_4$, $c_5$,
$c_6$, $c_7$, and $c_8$ survive and thus these coefficients must 
vanish.
\qedbox

\section{Manifolds with boundary}\label{SectBoundary}

\medbreak Let $M_0$ be a smooth Riemannian manifold with smooth 
boundary
$\partial M_0$ and let $D_0$ be an operator of Laplace type over $M_0$. 
Let
$$\mathcal{B}_D\phi:=\phi|_{\partial M_0}\text{ and }
  \mathcal{B}_S\phi:=(\nabla_\nu\phi+S\phi)|_{\partial M_0}.$$
The operator $\mathcal{B}_D$ defines Dirichlet boundary conditions and 
the
operator $\mathcal{B}_S$ defines Robin boundary conditions. Let
$D_{\mathcal{B}}$ be the realization of $D$ with the associated 
boundary
condition. If
$f$ is a smooth function on
$M$, then
\begin{eqnarray*}
  &&\trace_{L^2}(fe^{-tD_{{\mathcal{B}}_{D/S}}})\sim
  \textstyle\sum_{n\ge0}t^{(n-m)/2}a_n(f,D,{\mathcal{B}}_{D/S})\text{ 
where}\\
  &&a_n(f,D,{\mathcal{B}}_{D/S}) =a_n^M(f,D)+a_n^{\partial
M_0}(f,D,{\mathcal{B}}_{D/S})\end{eqnarray*} 
are given by local formulas. The interior
invariants $a_n^M(f,D)$ can be calculated using Theorem 
\ref{Theoremanm}.
Formulas if $n\le5$ are known for the invariants
$a_n^M(f,D,{\mathcal{B}}_{D/S})$ for
$n\le5$; see for example
\cite{BG90, BGV95, BGKV99, KCD80, K90, Moss00, MD,Vassilevich:1995we}. 
These results yield the
following:
\begin{lemma}\label{drefa}\ \begin{enumerate}
\item $a_0^{\partial M}(f,D,{\mathcal{B}}_{D/S})=0$.
\item $a_1^{\partial
M}(f,D,\mathcal{B}_D)=-(4\pi)^{(1-m)/2}\frac14\int_{\partial 
M}\trace(I)$.
\item $a_1^{\partial
M}(f,D,\mathcal{B}_S)=(4\pi)^{(1-m)/2}\frac14\int_{\partial 
M}\trace(I)$.
\item $a_2^{\partial 
M}(f,D,\mathcal{B}_D)=(4\pi)^{-m/2}\frac16\int_{\partial
M}
      \trace\{2fL_{aa}I-3f_{;m}I\}$.
\item $a_2^{\partial M}(f,D,\mathcal{B}_S)
      =(4\pi)^{-m/2}\frac16\int_{\partial
M}\trace\{f(2L_{aa}I+12S)+3f_{;m}I\}$.
\item $a_3^{\partial M}(f,D,\mathcal{B}_D)=-(4\pi)^{(1-m)/2}\frac1{384}
      \int_{\partial M}\trace\{96fE+f(16R_{ijji}$
\bork
$-8R_{amma}+7L_{aa}L_{bb}-10L_{ab}L_{ab})I-30f_{;m}L_{aa}I+24f_{;mm}I\}$.
\item $a_3^{\partial M}(f,D,\mathcal{B}_S)=+(4\pi)^{(1-m)/2}\frac1{384}
     \int_{\partial M}\trace(96fE+f(16R_{ijji}-8R_{amma}$
\bork$
     +13L_{aa}L_{bb}
     +2L_{ab}L_{ab})I+f(96SL_{aa}+192S^2)
     +f_{;m}(6L_{aa}I+96S)$
\bork$+24f_{;mm}I\}$.
\smallbreak\item $a_4^{\partial 
M}(f,D,\mathcal{B}_D)=(4\pi)^{-m/2}\frac1{360}
     \int_{\partial M}\trace\{f(-120E_{;m}+120EL_{aa})$
\bork
     $+f(-18R_{ijji;m}+20R_{ijji}L_{aa}
     +4R_{amam}L_{bb}
     -12R_{ambm}L_{ab}+4R_{abcb}L_{ac}$ 
     \bork
     $+24L_{aa:bb}+\frac{40}{21}L_{aa}L_{bb}L_{cc}
     -\frac{88}7L_{ab}L_{ab}L_{cc}+\frac{320}{21}L_{ab}L_{bc}L_{ac})I-180f_{;m}E
     $\bork$+f_{;m}(-30R_{ijji}
     -\frac{180}{7}L_{aa}L_{bb}+\frac{60}7L_{ab}L_{ab})I
     +24f_{;mm}L_{aa}I-30f_{;iim}I\}$.
\smallbreak\item $a_4^{\partial M}(f,D,\mathcal{B}_S)=
      (4\pi)^{-m/2}\frac1{360}\int_{\partial M}\
      \trace\{f(240E_{;m}+120EL_{aa})+f(42
R_{ijji;m}
     $
\bork$+24L_{aa:bb}+20 R_{ijji}L_{aa}+4R_{amam}L_{bb}
-12R_{ambm}L_{ab}+4R_{abcb}L_{ac}$
\bork$
+\frac{40}3L_{aa}L_{bb}L_{cc}
+8L_{ab}L_{ab}L_{cc}+\frac{32}3L_{ab}L_{bc}L_{ac})I+f(720SE+120S 
R_{ijji}$
\bork
$
+144SL_{aa}L_{bb}+48SL_{ab}L_{ab}+480S^2L_{aa}+480S^3+120S_{:aa})$
\bork
$
+f_{;m}(180E+72SL_{aa}+240S^2)+f_{;m}(30R_{ijji}+12L_{aa}L_{bb}+12L_{ab}L_{ab}
)I$\bork
$+120f_{;mm}S+24f_{;mm}L_{aa}I+30f_{;iim}I\}$.
\end{enumerate}\end{lemma}

We extend results of \cite{BV99} for smooth metrics to the
current setting to 
relate the invariants $a_n^{\partial M}(f,D,{\mathcal{B}}_{D/S})$ to 
the
invariants $a_n^\Sigma(f,D)$ as follows. 

\begin{lemma}\label{drefb} Let $M^\pm$ be two copies of of a smooth 
manifold
$M_0$ joined along the common boundary. Let $D^\pm:=D_0$ and let 
$\dvxi=-2S$.
Extend $f_0\in C^\infty(M_0)$ to $M$ as an even function $f$. Then
$$a_n^\Sigma(f,D,\dvxi)=a_n^{\partial M_0}(f_0,D_0,{\mathcal{B}}_D)+
  a_n^{\partial M_0}(f_0,D_0,{\mathcal{B}}_S).$$
\end{lemma}

\noindent{\bf Proof:} Let  $\{\lambda_{D,i},\tilde\phi_{D,i}\}$ and
$\{\lambda_{S,i},\tilde\phi_{S,i}\}$ be the discrete spectral 
resolutions of
$D_0$ for Dirichlet and Robin boundary conditions over $M_0$. We wish 
to
use these collections to construct the discrete spectral resolution of
$D$ with the given transmittal boundary conditions.
Extend the sections
$\tilde\phi_{D,i}$ to be odd sections and the $\tilde\phi_{S,i}$ to be 
even
sections on
$V$ over $M$ by defining:
$$\textstyle\phi_{D,i}(x^\pm)=\pm\frac1{\sqrt{2}}\tilde\phi_{D,i}(x)\text{ 
and }
  \phi_{S,i}(x^\pm)=\frac1{\sqrt{2}}\tilde\phi_{S,i}(x).$$
We show the sections $\{\phi_{D,i},\phi_{S,j}\}$ form an orthonormal
system by computing:
\begin{eqnarray*}
  &&\textstyle\int_M(\phi_{D,i},\phi_{D,j})=
    \textstyle\frac12\int_{M^+}(\tilde\phi_{D,i},\tilde\phi_{D,j})
  +(-1)^2\frac12\int_{M^-}(\tilde\phi_{D,i},\tilde\phi_{D,j})=\delta_{ij},\\
  &&\textstyle\int_M(\phi_{D,i},\phi_{S,j})
   =\frac12\int_{M^+}(\tilde\phi_{D,i},\tilde\phi_{S,j})
    -\frac12\int_{M^-}(\tilde\phi_{D,i},\tilde\phi_{S,j})=0,\text{ 
and}\\
  &&\textstyle\int_M(\phi_{S,i},\phi_{S,j})=
    \textstyle\frac12\int_{M^+}(\tilde\phi_{S,i},\tilde\phi_{S,j})
  +\frac12\int_{M^-}(\tilde\phi_{S,i},\tilde\phi_{S,j})=\delta_{ij}.
\end{eqnarray*}
Let $\phi=(\phi^+,\phi^-)$. We define:
$$\textstyle\tilde\phi_e(x)=\frac12(\phi^+(x^+)+\phi^-(x^-))\text{ and 
}
  \tilde\phi_o(x)=\frac12(\phi^+(x^+)-\phi^-(x^-)).$$
We expand $\tilde\phi_{e/o}$ using the sections $\phi_{S/D,j}$. We then
extend $\tilde\phi_{e/o}$ to an even/odd pair of sections $\phi_{e/o}$. 
Since
$\phi=\phi_e+\phi_o$, this shows that the sections
$\{\tilde\phi_{D,i},\tilde\phi_{S,i}\}$ form a complete orthonormal 
basis for
$L^2(V)$. Since the $\phi_{S,i}$ are even sections, they are 
continuous.
Since the $\phi_{D,i}$ are odd sections which vanish on the common 
boundary,
they are continuous as well. We verify that the transmittal boundary
conditions are satisfied by computing:
\begin{eqnarray*}
&&(\phi_{D,i;\nu^+}^+)|_\Sigma
      =\textstyle\frac1{\sqrt2}(\tilde\phi_{D,i;\nu})|_\Sigma
      =(-\phi_{D,i;\nu^-})|_\Sigma+0
      =(\phi_{D,i;\nu^+}^-)|_\Sigma-2S\phi_{D,i}|_\Sigma\\
&&(\phi_{S,i;\nu^+}^+)|_\Sigma
      =\textstyle\frac1{\sqrt2}(\tilde\phi_{S,i;\nu})|_\Sigma
      =\textstyle\frac1{\sqrt2}(\tilde\phi_{S,i;\nu})|_\Sigma-2
       \textstyle\frac1{\sqrt2}(\tilde\phi_{S,i;\nu}+
       S\tilde\phi_{S,i})|_\Sigma\\
&&\qquad\qquad=-\textstyle\frac1{\sqrt2}(\tilde\psi_{S,i;\nu})|_\Sigma
    -2S\psi_{S,i}|_\Sigma
     =-(\psi_{S,i;\nu^-}^-)|_\Sigma-2S\psi_{S,i}|_\Sigma\\
&&\qquad\qquad=(\psi_{S,i;\nu^+}^-)|_\Sigma-2S\psi_{S,i}|_\Sigma.
\end{eqnarray*}
This shows that
$\{(\lambda_{i,D},\phi_{i,D}),(\lambda_{j,S},\phi_{j,S})\}$ gives the 
desired
discrete spectral resolution. Since $f$ is an even function, we use 
equation
(\ref{arefab}) to see:
\begin{eqnarray*}
\trace_{L^2}(fe^{-tD})&=&
   \textstyle\sum_ie^{-t\lambda_{i,D}}\int_Mf(\phi_{i,D},\phi_{i,D})
   +\textstyle\sum_je^{-t\lambda_{j,S}}\int_Mf(\phi_{j,S},\phi_{j,S})\\
&=&\textstyle\sum_ie^{-t\lambda_{i,D}}\int_{M_0}
       f_0(\tilde\phi_{i,D},\tilde\phi_{i,D})
   +\textstyle\sum_je^{-t\lambda_{j,S}}\int_{M_0}
       f_0(\tilde\phi_{j,S},\tilde\phi_{j,S})\\
&=&\trace_{L^2}(f_0e^{-tD_{0,\mathcal{B}_D}})
   +\trace_{L^2}(f_0e^{-tD_{0,\mathcal{B}_S}}).
\end{eqnarray*} The desired result now follows by equating terms in the
asymptotic expansions. We note that were we to extend $f_0$ as an {\it 
odd}
function, then the terms would cancel instead of combining and we would 
get 0.
\qedbox

The following result is an immediate consequence of Lemmas
\ref{drefa} and \ref{drefb}.

\begin{lemma}\label{drefc} We have
$d_1=2$, $d_2=0$, $d_3=-6$, $d_4=\frac32$, $d_5=3$,
$d_6=9$, $d_7=48$, $d_8=-24$, and $d_9=-24$.
\end{lemma}

\section{A Transmittal problem for the de Rham cplx}\label{SectDeRham}

To evaluate the coefficient $e_1$, we use the local index theorem. We 
begin
by constructing transmittal boundary conditions for the de Rham 
complex. We begin
by recalling some facts concerning exterior $\pext$, interior $\pint$, 
and
Clifford multiplication $\cl$.  Let $\dvxi$ be a cotangent vector. We 
can choose a
local orthonormal frame so $\dvxi=ce^1$. Let $1\le i_1<...<i_p\le m$. 
Then
\begin{eqnarray*}
  &&\pext(\dvxi)e^{i_1}\wedge...\wedge e^{i_p}=
  ce^1\wedge e^{i_1}\wedge...\wedge e^{i_p}\text{ if }1<i_1,\\
  &&\pext(\dvxi)e^{i_1}\wedge...\wedge e^{i_p}=0\text{ if }1=i_1,\\
  &&\pint(\dvxi)e^{i_1}\wedge...\wedge e^{i_p}=0\text{ if }1<i_1,\text{ 
and}\\
  &&\pint(\dvxi)e^{i_1}\wedge...\wedge e^{i_p}=
  ce^{i_2}\wedge...\wedge e^{i_p}\text{ if }1=i_1.\end{eqnarray*}
Thus exterior multiplication adds an index and interior
multiplication cancels an index if possible. 
Let $\cl(\dvxi):=\pext(\dvxi)-\pint(\dvxi)$ denote Clifford 
multiplication;
$$\cl(\dvxi_1)\cl(\dvxi_2)+\cl(\dvxi_2)\cl(\dvxi_1)=-2(\dvxi_1,\dvxi_2)\id.$$
We can write the exterior derivative $d$ and the interior derivative 
$\delta$ in
the form:
\begin{eqnarray}
  &&d\phi^\pm=\pext(e^i)\nabla^\pm_{e_i}\phi^\pm,\qquad
  \delta\phi^\pm=-\pint(e^i)\nabla^\pm_{e_i}\phi^\pm,\text{ and}
\label{erefaA}\\
&&(d+\delta)\phi^\pm=\cl(e^i)\nabla^\pm_{e_i}\phi^\pm.\nonumber\end{eqnarray}

Let  $M=M^+\cup_\Sigma M^-$; we give $M$ the smooth structure defined 
in Section
\ref{SectSmooth}. Let $V:=\Lambda$ be the exterior algebra. Let
$\phi=(\phi^+,\phi^-)$ where $\phi^\pm$ are smooth differential forms 
over
$M^\pm$. We let $\phi^\pm|_\Sigma$ be sections to the full exterior 
bundle; we
do not set $dx^m$ to zero. Let
$${\mathcal{B}}_0\phi:=\phi^+|_\Sigma-\phi^-|_\Sigma.$$
Thus ${\mathcal{B}}_0\phi=0$ if and only if $\phi$ is continuous on 
$\Sigma$.
We let $\Delta:=(d+\delta)^2$ with
$$\text{Domain}(\Delta):={\mathcal{D}}:=\{\phi:{\mathcal{B}}_0\phi=0\text{ 
and }
   {\mathcal{B}}_0(d+\delta)\phi=0\}.$$

The following Lemma will be crucial for our analysis. 
\begin{lemma}\label{erefa}\ \begin{enumerate}
\item We have ${\mathcal{B}}_0\phi=0$ if and
only if
$((d+\delta)\phi,\psi)_{L^2}=(\phi,(d+\delta)\psi)_{L^2}$ for every 
$\psi$
satisfying ${\mathcal{B}}_0\psi=0$.
\item The operator $\Delta:=(d+\delta)^2$ with the domain $\mathcal{D}$ 
is
self-adjoint.
\item Let $\dvxi:=\cl(e^m)\cl(e^a)\omega_a$. Then  
$\phi\in{\mathcal{D}}$ if and
only if ${\mathcal{B}}_\dvxi\phi=0$.
\item Let $\LL_{ab}:=(L_{ab}^++L_{ab}^-)$. Then
$\omega_a=\LL_{ab}(\pext(e^m)\pint(e^b)+\pint(e^m)\pext(e^b))$ and
\phantom{.}\hfill$\dvxi=\LL_{ab}\{\pext(e^m)\pint(e^m)\pint(e^a)\pext(e^b)
 +\pint(e^m)\pext(e^m)\pext(e^a)\pint(e^b)\}$. 
\item Since
$\dvxi\Lambda^p\subset\Lambda^p$, $\dvxi$ induces transmittal boundary
conditions for $\Delta_p$. We have
$a_n(1,\Delta_e,\dvxi_e)-a_m(1,\Delta_o,\dvxi_0)=0$ for $n\ne m$ 
and\bork
$a_m(1,\Delta_e,\dvxi_e)-a_m(1,\Delta_o,\dvxi_o)\in\intgs$.
\item We have
$\dim\ker(\Delta^e_{\BB_\dvxi})-\dim\ker(\Delta^o_{\BB_\dvxi})=\chi(M)$.
\end{enumerate}\end{lemma}

\noindent{\bf Proof:} We apply equation (\ref{erefaA}) to study
$((d+\delta)\phi,\psi)_{L^2}-(\phi,(d+\delta)\psi)_{L^2}$. We can 
integrate by
parts to exchange tangential derivatives; these cancel automatically. 
We assume
$\psi^+|_\Sigma=\psi^-|_\Sigma$. After taking into account the 
different signs of
the relevant normals, Green's formula yields, modulo a possible sign 
convention
that plays no role,
\begin{eqnarray}
&&((d+\delta)\phi,\psi)_{L^2}-(\phi,(d+\delta)\psi)_{L^2}=
\textstyle\int_\Sigma\{\phi^+|_\Sigma
     -\phi^-|_\Sigma\}\cdot\{\cl(\nu)\psi|_\Sigma\}.\label{erefaa}
\end{eqnarray}
Since $(\cl{\nu})^2=-1$, the terms in equation
(\ref{erefaa}) vanish for all suitable $\psi$ if and only if
$\phi^+|_\Sigma=\phi^-|_\Sigma$.
If $\phi,\psi\in{\mathcal{D}}$, then we can use assertion (1) to show
that $D$ with this realization is self-adjoint by observing that:
\begin{eqnarray*}
&&(\Delta\phi,\psi)_{L^2}=((d+\delta)\phi,(d+\delta)\psi)_{L^2}\text{ 
since }
{\mathcal{B}}_0(d+\delta)\phi=0\text{ and }{\mathcal{B}}_0\psi=0\\
&&(\phi,\Delta\psi)_{L^2}=((d+\delta)\phi,(d+\delta)\psi)_{L^2}\text{ 
since }
{\mathcal{B}}_0\phi=0\text{ and }{\mathcal{B}}_0(d+\delta)\psi=0.
\end{eqnarray*}

Let ${\mathcal{B}}_0\phi=0$. We compute:
\begin{eqnarray}
&&\{(d+\delta^+)\phi^+\}|_\Sigma-\{(d+\delta^-)\phi^-\}|_\Sigma
   \label{erefab}\\
&=&\cl(e^i)\{(\nabla_i^+\phi^+)|_\Sigma-(\nabla_i^-\phi^-)|_\Sigma\}\nonumber\\
&=&\cl(e^a)\{\nabla_a^+-\nabla_a^-\}\phi|_\Sigma+\cl(e^m)
\{(\nabla_\nu^+\phi^+)|_\Sigma-(\nabla_\nu^-\phi^-)|_\Sigma\}\nonumber\\
&=&\cl(e^m)\{(\nabla_\nu^+\phi^+)|_\Sigma-(\nabla_\nu^-\phi^-)|_\Sigma
       -\cl(e^m)\cl(e^a)\omega_a\phi|_\Sigma\}.\nonumber
\end{eqnarray}
Since $\cl(e^m)$ is an isomorphism, the terms in (\ref{erefab}) vanish 
if and only if $\phi$ satisfies the transmittal boundary condition 
defined by
$\dvxi$.
We compute:
\begin{eqnarray*}
 &&\nabla_i(e^{i_1}\wedge...\wedge e^{i_p})=
   \textstyle\sum_{1\le j\le p}(-1)^{j-1}\Gamma_{ii_j\ell}
    e^{i_1}\wedge...\wedge e^\ell\wedge...\wedge e^{i_p}\\
 &&\qquad\qquad=\Gamma_{ik\ell}\pext(e^\ell)\pint(e^k)\\
 &&\omega_a=\LL_{ab}\{\pext(e^m)\pint(e^b)-\pext(e^b)\pint(e^m)\}\\
 &&=\LL_{ab}\{\pext(e^m)\pint(e^b)+\pint(e^m)\pext(e^b)\}
\end{eqnarray*}
If $i\ne j$, then
\begin{eqnarray*}
 &&\pext(e^i)\pext(e^i)+\pext(e^j)\pext(e^j)=0,\\ 
 &&\pint(e^i)\pext(e^j)+\pext(e^j)\pint(e^i),\text{ and}\\
 &&\pint(e^i)\pint(e^j)+\pint(e^j)\pint(e^j)=0.
\end{eqnarray*}
Furthermore $\pext(e^i)\pext(e^i)=0$ and $\pint(e^i)\pint(e^i)=0$. 
Consequently:
\begin{eqnarray*}
 \dvxi&=&\LL_{ab}(\pext(e^m)-\pint(e^m))(\pext(e^a)-\pint(e^a))\\
     &&\cdot(\pext(e^m)\pint(e^b)+\pint(e^m)\pext(e^b)\}\\
&=&\LL_{ab}\{\pext(e^m)\pint(e^m)\pint(e^a)\pext(e^b)\\
    &&\qquad+\pint(e^m)\pext(e^m)\pext(e^a)\pint(e^b)\}.
\end{eqnarray*}

We use assertion (4) to see that $\dvxi\Lambda^p\subset\Lambda^p$ since 
there are
two $\pext$ and two $\pint$ terms. It is also clear that $\dvxi$ is 
self-adjoint;
this gives another proof of assertion (2). We extend the cancellation 
argument of
McKean and Singer \cite{MS67} to prove assertion (5). Let 
$E(\lambda,\Delta,\dvxi)$
be the associated eigenspaces. Suppose $\Delta\phi=\lambda\phi$ and 
that
${\mathcal{B}}_\dvxi\phi=0$. Since $(d+\delta)\Delta=\Delta(d+\delta)$, 
we have
$\Delta(d+\delta)\phi=\lambda(d+\delta)\phi$. We show
that $(d+\delta)\phi\in{\mathcal{D}}$ by computing:
$${\mathcal{B}}_0(d+\delta)\phi=0\text{ and }
{\mathcal{B}}_0(d+\delta)(d+\delta)\phi=\lambda{\mathcal{B}}_0\phi=0.$$
Thus we have $(d+\delta):E(\lambda,\Delta_{e/o},\dvxi_{e/o})\rightarrow
    E(\lambda,\Delta_{o/e},\dvxi_{o/e})$.
If $\lambda\ne0$, then $(d+\delta)^2=\lambda$ is an isomorphism and 
thus
\begin{equation}
   \dim(E(\lambda,D_e,\dvxi_e))-\dim(E(\lambda,D_o,\dvxi_o))=0\text{
for }
  \lambda\ne0.\label{erefac}\end{equation}
We use equation (\ref{erefac}) to compute:
\begin{eqnarray}
&&\trace_{L^2}(e^{-tD_\dvxi^e})-\trace_{L^2}(e^{-tD_\dvxi^o})\label{erefad}\\
&=&\textstyle\sum_\lambda 
e^{-t\lambda}\{\dim(E(\lambda,\Delta_e,\dvxi_e))
   -\dim(E(\lambda,\Delta_o,\dvxi_o))\}\nonumber\\
&=&\dim(E(0,\Delta_e,\dvxi_e))-\dim(E(0,\Delta_o,\dvxi_o)).\label{erefae}
\end{eqnarray}
We compare coefficients of powers of $t$ in the asymptotic expansion on 
the
left in (\ref{erefad}) with (\ref{erefae}) to see that the constant 
term is an
integer and the other terms vanish.

It now follows that the index is given by a local formula and thus is 
constant
under deformations. We fix the metric on
$\Sigma$ and deform the metrics on $M^\pm$ so that the metric is 
product near
the boundary. We then have $\dvxi=0$ and $\Sigma$ no longer plays a 
role. Thus the
index is given by $\int_M\{a_m^M(\Delta^e)-a_m^M(\Delta^o)\}$ and the 
standard
local index theorem shows the index to be the Euler-Poincare 
characteristic
$\chi(M)$.
\qedbox

\medbreak We can now determine the remaining unknown coefficient:

\begin{lemma}\label{{erefd}} We have $e_1=24$.\end{lemma}

\medbreak\noindent{\bf Proof:} We apply Lemma \ref{erefa} with $m=2$.
Invariants which are multiplied by $\trace(I)$ cancel in the 
alternating sum.
We set
$f=1$ so the derivatives of $f$ play no role. Thus the only terms which
survive involve
$\dvxi$ and $\omega_a$ on $\Lambda^1$. Let $\LL:=\LL_{11}$.
We use Lemma \ref{erefa} to see that:
\begin{eqnarray*}
  &&\omega_1(1)=0\nonumber,\ \omega_1(e^1\wedge e^2)=0,\ 
    \omega_1(e^1)=\LL e^2,\ 
    \omega_1(e^2)=-\LL e^1,\\
&&\dvxi_0=0,\ \dvxi_2=0,\ 
  \dvxi_1(e^1)=\LL e^1,\ \dvxi_1(e^2)=\LL e^2,\\
&&(L_{aa}^++L_{aa}^-)\trace(\dvxi)=2\LL^2,\ 
  \trace(\dvxi^2)=2\LL^2,\text{ and }
  \trace(\omega_a\omega_a)=-2\LL^2\end{eqnarray*}
We use Lemma \ref{erefa} (5) to see that:
$$\textstyle\int_\Sigma\trace(e_1\omega_1\omega_1
  +48\dvxi_1^2-24(L_{11}^++L_{11}^-)\dvxi_1)=0.$$
Consequently $-2e_1+96-48=0$ so $e_1=24$. \qedbox

\section{Computation of the fourth order invariant}\label{SectA4}
In this section, we study the fourth order invariant. Before writing 
down a
spanning set for the space of invariants, we make the following 
observations.

Certain invariants have been omitted because they are chiral - i.e. 
they change
sign if we interchange the roles of $+$ and $-$. We omit these 
invariants -
typical examples would be
$$\trace\{f\dvxi\omega_{a:a},\ f\laas\omega_{b:b}\}.$$
If
$V^\pm$ are real vector bundles and if the operators $D^\pm$ are real 
operators,
then the invariants are real. Consequently, the coefficients are real. 
We
suppose given fiber metrics on the bundles $V^\pm$ and we suppose that 
the
operators $D^\pm$ are formally self-adjoint. Let $\dvxi$ be 
self-adjoint. The
calculations of equation (\ref{brefaa}) then show
$D$ is self-adjoint so again the invariants are real. Since 
$\trace(\Omega)$ and
$\trace(\omega_a)$ are purely imaginary if $D$ is self-adjoint, this 
observation
shows that the following invariants play no role in the computation of 
$a_4$:
\begin{eqnarray*}
  &&\{f\trace(\Omega^+_{a\nu^+;a}+\Omega^-_{a\nu^-;a}),\ 
    f(L_{bb}^+-L_{bb}^-)\trace(\omega_{a;a}),\ 
    f(L_{ab}^+-L_{ab}^-)\trace(\omega_{a;b}),\\
&&\phantom{\{}f(L_{ab:b}^+-L_{ab:b}^-)\trace(\omega_a),\ 
  f(L_{bb:a}^+-L_{bb:a}^-)\trace(\omega_a),\ 
  f(\rho_{am}^+-\rho_{am}^-)\trace(\omega_a),\\
&&\phantom{\{}f(L_{bb:a}^+-L_{bb:a}^-)\trace(\omega_a),\ 
   f(L_{ab:b}^+-L_{ab:b}^-)\trace(\omega_a)\}.\end{eqnarray*}

We can formulate now the main result of this section.

\begin{theorem}\label{w01a4}
1. There exist universal constants ${\bf b}=(b_1,\dots ,b_{20})$
such that
$$
a_4^\Sigma (f,D,\dvxi )=(4\pi)^{-m/2} 360^{-1} \textstyle\int_\Sigma
{\rm Tr} \left(
{\cal A}_1 +{\cal A}_2 +{\cal A}_3\right)
$$
where
\begin{eqnarray}
&&{\cal A}_1=  b_{1} \ea \fma +
b_{2} \ra \fma \nn \\
& &\quad +b_{3} \rmma \fma +b_{4} f \laaa \lbba \nn\\ & &\quad 
\times\lccs 
+b_{5} f \laba \laba \lccs +b_{6} f \labs\nn\\ & & \quad\times \laba 
\lcca 
+ b_{7} f \laba \lbca \lcas \nn\\ & &\quad+ b_{8} \laaa \lbba \fms 
 + b_{9} \laba \laba \nn\\ & &\quad \times \fms + b_{10} \laaa \lbbs 
\fma \nn\\
& &\quad +b_{11} \laba \labs  \fma 
+ b_{12} f \ea \laaa \nn\\
& &\quad +b_{13} f \ra \laaa 
+ b_{14} f \rmma \nn\\ & &\quad \times  \lbba + b_{15} f \rambma \laba 
\nn\\
& &\quad + b_{16} f \rabcba \laca + b_{17} \laaa \fmma \nn\\
& &\quad + b_{18} \omega_a^2 \fms +b_{19} f \omega_a^2 \lbbs 
+b_{20} f \omega_a \omega_b \labs \nn\\
&&{\cal A}_2=  
60 f \ems 
+12 f \rms + 0 \es \fms \nn\\
& &\quad + 0 \rs \fms + 0 \rmms \fms \nn\\
& &\quad + 0 \left[(\Delta f)_{;\nu^+} + (\Delta f)_{;\nu^-}\right] 
-60 f \omega_a \left( \Omega_{a\nu^+} - \Omega_{a\nu^-}\right) \nn\\
& &\quad + \frac{40}{21} f \laas \lbbs \lccs -\frac 47 f \labs 
\nn\\ & &\quad \times \labs \lccs 
+ \frac{68}{21} f \labs \lbcs \lcas \nn\\
& &\quad -\frac{12}7 \laas \lbbs \fms 
+\frac {18}7 \labs \nn\\
& &\quad \times\labs \fms + 24 f (L^+ _{aa:bb} + L^-_{aa:bb}) \nn\\
& &\quad + 0 f (L^+ _{ab:ab} + L^- _{ab:ab}) 
+ 60 f \es \laas \nn\\
& &\quad + 10 f \rs \laas + 2 f \rmms \laas \nn\\
& &\quad - 6 f \rambms \labs + 2 f \rabcbs \lacs \nn\\
& &\quad + 12 \laas \fmms \nn\\
&&{\cal A}_3=
-60 f \dvxi^3 - 30 f \dvxi \rs 
- 180 f \dvxi \es - 60 f \dvxi_{:aa} \nn\\
& &\quad + 0 f \dvxi \rmms + 15 \dvxi \laaa \fma \nn\\
& &\quad - 9 \dvxi \laas  \fms + 0 f \dvxi \laaa \lbba \nn\\
& &\quad + 0 f \dvxi \laba \laba - 18 f \dvxi \laas \lbbs \nn\\
& &\quad  -6 f \dvxi \labs \labs - 30 \dvxi \fmms \nn\\
& &\quad + 30 \dvxi^2 \fms + 60 f \dvxi^2 \laas 
-60 f \dvxi \omega_a ^2   \nn
\end{eqnarray}
2. The universal constants are given by

\begin{tabular}{|l|l|l|l|l|}\hline
$b_{1}=-30$ & $b_{2}=-5$ & $b_{3}=2$ & $b_{4}=0$ & 
$b_{5}=-1$ \\ \hline $b_{6}=-1$ &
 $b_{7}=2$ & $b_{8} = 0 $ & 
$b_{9}=0$ & $b_{10}=-5$ \\ \hline $b_{11}=-1$ & $b_{12}=0$ & 
$b_{13}=0$ & $b_{14}=0$ & $b_{15}=0$ \\ \hline
$b_{16}=2$ & 
$b_{17}=0$ & $b_{18} = 18$ &
$b_{19} = 12$ & $b_{20}=24$ \\ \hline
\end{tabular}
\end{theorem}

\medbreak\noindent{\bf Proof:} We can use the analysis of section \ref{SectInvar} and
the list of invariants which cannot contribute to $a_4^\Sigma$ which was given at the
beginning of this section to determine the general form of the invariant $a_4^\Sigma$.
Coefficients of the invariants contained in
${\cal  A}_3$
have been calculated in \cite{BV99,Moss00}. The coefficients listed
in ${\cal A}_2$ follow from Lemma \ref{drefb} and the heat trace
asymptotics for Dirichlet and Neumann boundary conditions 
(see Lemma \ref{drefa}).

Next we use the conformal properties of the heat kernel
coefficients\footnote{Note that we define the conformal transformations
in such a way to make $D$ conformally covariant. These transformations
do not necessarily coincide with the conformal (Weyl) transformations
adopted in physics.}
\cite{BG90}.

\begin{lemma}\label{w01conf}
Let $ D( \epsilon )=e^{ -2 \epsilon
f}D$. We then have that $$  \birdy 
a_n(1,D(\epsilon))=(m-n)a_n(f,D(0)).$$
\end{lemma}

We suppose that the conformal transformation parameter
$f$ is continuous but not necessarily smooth across $\Sigma$.
The metric transforms as $g(\epsilon)=e^{2f\epsilon}g$.
We have the following relations; a more extensive list
is given in \cite{BG90,BGKV99}, conformal variations
of all invariants relevant for calculation of $a_4^\Sigma$
are listed in Appendix:
\begin{eqnarray}
&&\birdy L^\pm_{ab} =-fL^\pm_{ab} -f_{;\nu^\pm}\delta_{ab}\,,\nn\\
&&\birdy E^\pm =-2fE^\pm +\frac 12 (m-2)f_{;\nu^\pm} \,,\nn\\
&&\birdy \dvxi =-f\dvxi -\frac 12 (m-2)\fms \,.\nn
\end{eqnarray}

We put $n=4$ in Lemma \ref{w01conf} and
 collect the terms with $(E^+-E^-)\fma$,
$\omega_a^2\fms$,  $\laaa\lbbs\fma$, 
$\ra\fma$, $\rmma\fma$, $\laaa\fmma$, 
$\laaa\lbba\fms$, $\laba\laba\fms$, and
$\labs\laba\fma$
to obtain, respectively,
\begin{eqnarray}
&&0=-2(m-1)b_{12} -60 (m-4)-2(m-4)b_{1} \,,\label{rel1}\\
&&0=-30 (2-m) -(m-1) b_{19} -b_{20} 
-(m-4) b_{18} \,,\label{rel2}\\
&&0=2(m-1) b_{4} + 2b_{5} +b_{6} +\frac 1 4 (m-2) b_{12} \nn \\
&&\quad -(m-1) b_{13} +\frac 1 2 b_{14} +\frac 1 2 b_{16} + 15
 (m-2) \nn  \\
&&\quad -10(m-1)  +2 +(m-4) b_{10}\,,
\label{rel3}\\
&&0=-5m+18 -(m-1)b_{13}+b_{16}-(m-4)b_2 \,,\label{rel4}\\
&&0= 2(m-6) -(m-1) b_{14} -b_{15} +2b_{16} -(m-4) b_{3}\,,
\label{rel5}\\
&&0=\frac 1 2 (m-2) b_{12} -2(m-1) b_{13} 
+(m-1) b_{14} +b_{15}
-(m-4) b_{17}\,,\label{rel6}\\
&&0=-(m-1) b_{4} -b_{6} -\frac 1 4 (m-2) b_{12} 
         +(m-1) b_{13} \nn \\
&&\quad -\frac 1 2 b_{14} -\frac 1 2 b_{16} 
-(m-4) b_{8} \,,\label{rel7} \\
&&0=-(m-1) b_{5} -b_{7} 
-\frac 1 2 b_{15} -\frac 1 2 (m-3)
      b_{16}  -(m-4) b_{9}\,,
\label{rel8}\\
&&0=-(m-1)b_6 -2b_7+3-\frac 12 b_{15}-(m-3) \nn\\
&&\quad -\frac 12 (m-3)b_{16} -2(m-4)b_{11}\,. \label{rel9}
\end{eqnarray}
Since the universal constants $b_i$ do not depend upon
dimensions $m$ we obtain from eq. (\ref{rel1}):
\begin{equation}
b_{12}=0,\qquad b_{1}=-30 \label{41241}
\end{equation}

Next we consider the de Rham complex and use Lemma \ref{erefa}
to calculate

\begin{lemma}\label{w01dera4}
$b_{19}+b_{20}=36$.\end{lemma}
\noindent
We omit the proof as it goes along exactly the same lines as that used to prove Lemma
\ref{{erefd}}.

Together with the equation (\ref{rel2}) above Lemma \ref{w01dera4}
gives:
\begin{equation}
b_{18}=18\,,\qquad b_{19}=12\,,\qquad b_{20}=24\,.
\label{181920}
\end{equation}

The last ingredient which we use in this section is the
following special case calculation:
\begin{lemma}\label{w01part}
Let $M^+$ be a unit hemisphere and $M^-$ be a unit ball.
Let $D^\pm$ be a scalar Laplacian with $E^+=-\frac 14 (m-1)^2$
and $E^-=0$. Let $\dvxi =(m-2)/2$. Then
$$
360 \Gamma (m/2) 2^{m-1} a_4^\Sigma (1,D,\dvxi)=
-\frac{250} 7 +\frac{2839} {42} m -\frac{191} 7 m^2 +\frac{61}{21} m^3 
.
$$
\end{lemma}
\noindent {\bf Proof}: By applying transmittal boundary conditions
to eigenfunctions of the Laplace operator on the hemisphere and
on the ball we obtain the following implicit 
equation for the eigenvalues $\lambda$:
\begin{equation}
\lambda J_{\m} ' (\lambda ) P_{\lambda -1/2} ^{-(\m)} (0) 
-J_{\m} (\lambda) \frac d {dx} P_{\lambda -1/2} ^{-(\m)} (x) |_{x=0} = 
0\,,
\label{spe1}
\end{equation}
where $J$ and $P$ are the Bessel and associated 
Legendre functions respectively.
The heat kernel asymptotics is now calculated by applying
the technique of \cite{Bordag:1996gm,Bordag:1996fw}. We do not give 
here details of this lengthy calculation.\qedbox

Lemma \ref{w01part} and equations (\ref{rel3}) - (\ref{rel9})
fix the remaining universal constants. This completes the proof
of Theorem \ref{w01a4}.\qedbox

Let us mention, that by exploiting all conformal relations
the numerical multipliers occuring in ${\cal A}_3$ have been 
fully confirmed.

\section{$a_5$ and renormalization of the brane-world scenario}
A complete calculation of the fifth order term is hardly
possible. Therefore, we restrict ourselves to a particular
case relevant for a discussion of the divergences in the
brane-world model. We suppose that the background field
configuration is approximately symmetric under reflection
about the surface $\Sigma$. This class of problems includes
the standard brane described by the metric (\ref{bwmet})
together with some reasonable generalizations.

\begin{lemma}\label{a5bw}
Let all left and right limits of all non-chiral invariants
up to dimension four coincide on the surface $\Sigma$ while
such limits of all chiral invariants up to dimension four
change sign. Then
\begin{eqnarray}
&&a_5(1,D,\dvxi )=\frac 1{5760} (4\pi )^{-(m-1)/2} 
\textstyle\int_\Sigma {\rm Tr}
\left\{ -720 E_{;\nu}\dvxi +120 \tau \dvxi^2 -135 \tau_{;\nu}\dvxi
\right. \nonumber \\
&&\qquad\qquad\quad 
 +30 \rho_{\nu\nu} \dvxi^2 + 240 \dvxi \dvxi_{:aa} + 720 E\dvxi^2 
+90\dvxi^4
 +450 \Omega_{a\nu} \Omega_{a\nu}
\nonumber \\
&&\qquad\qquad\quad
+ 540 L_{aa} E_{;\nu} +\frac{195}2 L_{aa}\tau_{;\nu} +30 L_{ab}
R_{a\nu\nu b;\nu} -135 L_{aa} \dvxi_{:bb}
\nonumber \\
&&\qquad\qquad\quad
-\frac{195}4 L_{aa:c}L_{bb:c} -\frac{75}2 L_{ab:c}L_{ab:c}
+30 L_{ab:a}L_{bc:c} -720 L_{aa}E\dvxi
\nonumber \\
&&\qquad\qquad\quad
-15 L_{aa}\dvxi \rho_{\nu\nu} -120 L_{aa}\dvxi \tau + 30 
L_{ab}\rho_{ab}\dvxi
-90 L_{ab}\dvxi R_{a\nu\nu b} 
\nonumber \\
&&\qquad\qquad\quad
+ 90 L_{aa}L_{bb} E +180 L_{ab}L_{ab} E + 15 L_{aa} L_{bb} \tau
+30 L_{ab}L_{ab} \tau
\nonumber \\
&&\qquad\qquad\quad
-\frac {15}4 L_{aa}L_{bb}\rho_{\nu\nu} -\frac {15}2 
L_{ab}L_{ab}\rho_{\nu\nu}
-15 L_{cc} L_{ab} \rho_{ab}
\nonumber \\
&&\qquad\qquad\quad
+ 15 L_{cc}L_{ab} R_{a\nu\nu b} -45 L_{ac}L_{ab}\rho_{bc}
+135 L_{ab} L_{ac} R_{b\nu\nu c}
\nonumber \\
&&\qquad\qquad\quad
- 45 L_{ac}L_{db} R_{dacb} -\frac{315}4 L_{cc}L_{ab}L_{ab}\dvxi
-75 L_{ab}L_{bc}L_{ac}\dvxi 
\nonumber \\
&&\qquad\qquad\quad
+270 L_{aa}L_{bb}\dvxi^2 + 90 L_{ab}L_{ab} \dvxi^2 -\frac {885}8
L_{aa}L_{bb}L_{cc}\dvxi -270 L_{aa}\dvxi^3
\nonumber \\
&&\qquad\qquad\quad
+\frac{1053}{64} L_{aa}L_{bb}L_{cc}L_{dd} +\frac{279}{16}
L_{cc}L_{dd}L_{ab}L_{ab} -\frac {921}{16} L_{ab}L_{ab}
L_{cd}L_{cd}
\nonumber \\
&&\qquad\qquad\quad
\left. -\frac{57}2 L_{dd}L_{ab}L_{bc}L_{ac} 
+\frac{639}4 L_{ab}L_{bc}L_{cd}L_{ad} \right\}
\label{a5}
\end{eqnarray}\end{lemma}

We recall the definition of chirality which was given
before Theorem \ref{ThmA4Sigma}.
Proof of this Lemma follows immediately from Lemma \ref{drefb}
and the expressions for the $a_5$ for Dirichlet and Robin
boundary value problem \cite{BGV95,K90,BGKV99}.

If the operator $D$ transforms covariantly under the Weyl
rescalings, the coefficient $a_5$ is proportional to the
Weyl anomaly and can be used to derive the corresponding
anomalous action.

The equation (\ref{a5})  represents the one-loop counterterms.
A theory is multiplicatively renormalizable only if all
independent counterterms are contained in the classical
action. Of course, we cannot expect that a theory containing
the Einstein gravity will be multiplicatively renormalizable.
One can hope, nevertheless, that renormalizability will be 
maintained at least
in the matter sector. This seems however not easy to achieve.
Consider for example the classical surface action of the
form
\begin{equation}
S_{cl}=\int_\Sigma d^4x\sqrt{h} W(\phi ) \label{Scl}
\end{equation}
Then the coefficient in front of the $\delta$-term in (\ref{singop})
is given by the second derivative of $W$ with respect to 
$\phi$, $\dvxi \propto W''(\phi )$. In the equation
(\ref{a5}) we see a term $\dvxi^4$. Such a term should be contained
in the classical action. This yields $(W''(\phi ))^4\propto W(\phi )$.
This last condition is satisfied by a rather exotic potential
$W(\phi )\propto \phi^{8/3}$. Phenomenological consequences
of such a potential are quite unclear.  

Certain simplifications could be achieved if one goes on shell,
i.e. if it is supposed that the background fields satisfy
their equations of motion.
This will
reduce the number of independent invariants. For example,
the extrinsic curvature $L_{ab}$ will be expressed by the
Israel conditions \cite{Israel66} through the surface stress-energy
tensor (which is essentially $g_{ab}\dvxi $ in the simplified
example (\ref{Scl})). Not much however can be gained on this way.
First, going on shell has nothing to do with strict renormalization
procedure of quantum field theory. Second, the number of divergent
terms will be still considerable. As usual, supersymmetry leads
to partial cancellation of the ultra violet divergences. For
example, all terms of purely geometrical origin (i.e. without $E$,
$\dvxi$ and $\Omega$) will go away just due to the balance of 
bosonic and fermionic degrees of freedom. For recent work on
supersymmetric brane-world scenario see 
\cite{Bagger:2000eh,Bergshoeff:2000ii}.

On the other hand, one may adopt a more radical and perhaps
more fruitful point of view borrowed from string models.
After separation of a few essential couplings, vanishing 
of the divergent field-dependent coefficients in front
of these couplings could be considered as a restriction
on the possible form of the (low-energy) background. Such
restrictions may play a role of equations of motion for
some effective theory. Practical realization of
this scenario is far from being clear.

Finally we stress that the heat trace asymptotics is local.
If there are more than one brane in the space-time the
coefficients $a_n$ are just sums of contributions of
individual branes.

\section{Acknowledgements}
We are grateful to M.~Bordag, S.~Dowker and D.~Fursaev for fruitful
discussions. 
Research of PBG partially supported by the NSF (USA)
and MPI (Leipzig, Germany).
 Research of KK partially supported by the MPI
(Leipzig, Germany) and the EPSRC under grant number GR/M45726.
 Research of DVV partially supported
by the DFG, project Bo 1112/11-1 (Germany) and ESI (Austria).

\section*{Appendix: Conformal variations}
Here we list the conformal variations which have been used
to obtain the relations (\ref{rel1})-(\ref{rel9}). We integrate by 
parts
where necessary to bring the variations into standard form so that
$f$ is not differentiated tangentially. 
We will be dealing with the terms $X$
which are homogeneous of dimension 3. If $f$ is constant, then
$\birdy X =-3fX$. To avoid writing the conformal weight repeatedly,
we define ${\cal C}X:=\birdy X +3fX$.
\begin{eqnarray}
b_4&&{\cal C}  \laaa \lbba \lccs = \nn \\
&&\qquad -2(m-1)\laaa \lbbs \fma \nn\\
&&\qquad -(m-1) \laaa \lbba \fms \nn \\ 
b_5&&{\cal C} \laba \laba \lccs = \nn \\
&&\qquad -2\laaa \lccs \fma \nn \\
&&\qquad-(m-1) \laba \laba \fms \nn \\
b_6&&{\cal C}  \labs \laba \lcca = \nn \\
&&\qquad -\laaa \lcca \fms \nn\\
&&\qquad -\laas \lcca \fma \nn\\
&&\qquad -(m-1)\labs \laba \fma \nn\\
b_7&&{\cal C}  \laba \lbca \lcas = \nn \\
&&\qquad -2\laba \labs \fma \nn\\
&&\qquad - \laba \laba \fms \nn \\
b_{12}&&{\cal C} \ea \laaa =
 -(m-1) \ea \fma \nn\\
&&\qquad +\frac 1 2 (m-2) \laaa \fmma \nn \\
&&\qquad -\frac 1 4 (m-2) \laaa \lbba \fms \nn\\
&&\qquad - \frac 1 4 (m-2) \laaa \lbbs \fma
\nn\\
b_{13}&&{\cal C} \ra \laaa =
-(m-1) \ra \fma \nn\\
&&\qquad -2(m-1) \laaa \fmma \nn\\
&&\qquad +(m-1) \laaa \lbba \fms \nn\\
&&\qquad +(m-1) \laaa \lbbs \fma 
\nn\\
b_{14}&&{\cal C} \rmma \lbba = 
(m-1) \laaa \fmma \nn\\
&&\qquad -\frac 1 2 \laaa \lbba \fms \nn\\
&&\qquad - \frac 1 2 \laas \lbba \fma \nn\\
&&\qquad - (m-1) \rmma \fma
\nn\\
b_{15}&&{\cal C} \rambma \laba =
-\rmma \fma \nn\\
&&\qquad  +\laaa \fmma \nn \\
&&\qquad -\frac 1 2 \laba \laba \fms \nn\\
&&\qquad -\frac 1 2 \laba \labs \fma
\nn\\
b_{16}&&{\cal C} \rabcba \laca =2 \rmma \fma \nn\\
&&\qquad -\frac 1 2 (m-3) \laba \laba \fms \nn \\
&&\qquad -\frac 1 2 (m-3) \laba \labs \fma \nn \\
&&\qquad -\frac 1 2 \laaa \lbba \fms \nn\\
&&\qquad -\frac 1 2 \laaa \lbbs \fma \nn \\
&&\qquad +\ra \fma 
\nn \\
b_{19}&&{\cal C} \omega_a^2 \lbbs =-(m-1) \omega_a ^2\fms \nn\\
b_{20}&&{\cal C} \omega_a \omega_b \labs =-\omega_a ^2 \fms \nn\\
60&&{\cal C} \es \laas=
-(m-1) \es \fms \nn\\
&&\qquad +\frac 1 2 (m-2) \laas \fmms  
+(m-2) f(L^+_{aa:bb} + L^- _{aa:bb}) \nn\\ 
&&\qquad -\frac 1 4 (m-2) \laas \lbbs \fms \nn \\
&&\qquad -\frac 1 4 (m-2) \laas \lbba \fma
 \nn\\
10&&{\cal C} \rs \laas =
-(m-1) \rs \fms \nn\\
&&\qquad -2(m-1) \laas \fmms
-4(m-1) f (L^+_{aa:bb} \nn\\
&&\qquad + L^-_{aa:bb}) +(m-1) \laas \lbbs \fms \nn \\
&&\qquad +(m-1) \laas \lbba \fma
 \nn\\
2&&{\cal C} \rmms \laas =
(m-1) \laas \fmms \nn\\
&&\qquad -\frac 1 2 \laas \lbbs \fms \nn \\
&&\qquad -\frac 1 2 \laas \lbba \fma+2f (L^+_{aa:bb} + L^- _{aa:bb}) 
\nn \\
&&\qquad -(m-1) \rmms \fms
\nn\\
-6&&{\cal C} \rambms \labs =
-\rmms \fms \nn\\
&&\qquad +\laas \fmms 
+2f(L^+_{ab:ab} \nn\\
&&\qquad + L^-_{ab:ab}) -\frac 1 2 \labs \labs \fms \nn \\
&&\qquad -\frac 1 2 \labs \laba \fma
 \nn\\
2&&{\cal C} \rabcbs \lacs =
 2f(m-3) (L^+_{ab:ab} + L^-_{ab:ab})\nn\\
&&\qquad  -\frac 1 2 (m-3) \labs 
                  \labs \fms\nn\\
&&\qquad -\frac 1 2 (m-3) \labs \laba \fma \nn\\
&&\qquad +2f (L^+_{aa:bb} + L^- _{aa:bb}) 
-\frac 1 2 \laas \lbbs \fms \nn\\
&&\qquad -\frac 1 2 \laas \lbba \fma \nn \\
&&\qquad +\rs \fms +2\rmms \fms \nn \\
-60&&{\cal C}\dvxi \omega_a^2 =-\frac 12 (m-2) \omega_a^2 \fms \nn
\end{eqnarray}
One should add the surface terms arising due to conformal variation
of the bulk terms $a^\pm (1,D)$ (see Theorem \ref{Theoremanm}).
 The total derivative terms $E_{;jj}$ and
$\tau_{;jj}$ in $a_4(1,D)$ cancel the surface terms with
$(E^+_{;\nu^+}+E^-_{;\nu^-})$ and $(\tau^+_{;\nu^+}+\tau^-_{;\nu^-})$.
Surface contributions of the conformal variation of the interior
terms in $M$ are:
\begin{eqnarray}
&&\frac 1{360} (4\pi )^{-m/2} {\textstyle\int_\Sigma} {\rm Tr} 
\{(12m-48) f(\tau_{;\nu^+}+\tau_{;\nu^-})
 +(-5m+18)(\fms \nn\\
&&\quad \times \rs 
  -\fma\ra ) 
 +60(m-4)
f\ems \nn\\
&&\quad -30(m-4)(\fms (E^++E^-)+\fma (E^+-E^-)) \nonumber \\
&&\quad +(2m-12)(\fms \rmms +\fma \nn\\
&&\quad \times\rmma ) 
 +(4n-24)f(L_{aa:bb}^+-L_{ab:ab}^++L_{aa:bb}^--L_{ab:ab}^-)
\} \nn
\end{eqnarray}

\end{document}